%% file: main.tex
\documentclass{article}
\usepackage{spconf,amsmath,graphicx}
\usepackage{subcaption}
\usepackage{booktabs}
\usepackage{makecell}
\usepackage[dvipsnames ]{xcolor}
\usepackage{tikz}
\usepackage{pgfplots}
\usepackage[numbers,square, sort]{natbib}
\usepackage{hyperref}
\usepackage{cleveref}

\newcommand\para[1]{\noindent \textbf{#1}}

\title{Contextual Speech Recognition with difficult\\negative training examples}
\name{}

\twoauthors{Uri Alon\sthanks{Work performed while at Google.}}
    {Technion, Israel\\
    \fontsize{9}{9}\selectfont\ttfamily\upshape
    \texttt{urialon@cs.technion.ac.il}}
    {Golan Pundak, Tara N. Sainath}
    {Google Inc., USA\\
    \fontsize{9}{9}\selectfont\ttfamily\upshape
    \texttt{\{golan,tsainath\}@google.com}}
\begin{document}
%
\maketitle
\input{abstract}

\input{intro}
\input{background}

\input{our_approach}
\input{experiments}
\input{qualitative}
\input{conclusion}

\section{Acknowledgements}

We would like to thank Rohit Prabhavalkar, Tony Bruguier, Alex Dovlecel, Dan Liebling, Jinxi Guo and Bo Li for their help.

\vfill\pagebreak

\bibliographystyle{IEEEbib}
\bibliography{strings,refs}

\end{document}

%% file: abstract.tex
\begin{abstract}

Improving the representation of contextual information is key to unlocking the potential of end-to-end (E2E) automatic speech recognition (ASR).
In this work, we present a novel and simple approach for training an ASR context mechanism with difficult negative examples. The main idea is to focus on proper nouns (e.g., unique entities such as names of people and places) in the reference transcript, and use phonetically similar phrases as negative examples, encouraging the neural model to learn more discriminative representations.
We apply our approach to an end-to-end contextual ASR model that jointly learns to transcribe and select the correct context items, and show that our proposed method gives up to $53.1\%$ relative improvement in word error rate (WER) across several benchmarks.
\end{abstract}

%% file: intro.tex
\section{Introduction}\label{sec:introduction}
End-to-end (E2E) models for ASR became popular in the last few years, as a way to replace the acoustic, language and pronunciation models of the ASR system with a single neural network \cite{Chan15,Sak15,SoltauLiaoSak16, BahdanauChorowskiSerdyukEtAl16, PrabhavalkarSainathWuEtAl18}. Recently, E2E models have achieved state-of-the-art performance on a Voice Search task \cite{CC18}. However, these models do poorly on rare and out-of-vocabulary (OoV) words. This issue is even more apparent when the system must recognize user-specific contextual information, which is an important component of a production-level ASR system \cite{AleksicAllauzenElsonEtAl15}. User context can include the user's favorite songs, contacts or apps, which often contain rare words.  

An effective remedy to this problem is to inject additional context into the system, to both inform the system on the possible presence of specific rare words as well as help it to identify more likely phrases. One possibility to introduce context to an E2E system is to interpolate the E2E model with an externally trained contextual language model \cite{WilliamsKannanAleksicEtAl18}, a process known as ``shallow fusion''. Alternatively, \cite{pundak2018deep} explores an all-neural approach for contextual biasing, by embedding a set of contextual phrases and allowing the decoder of the E2E model to attend to these contextual phrases. This approach was found to outperform shallow-fusion biasing on many tasks. However, if the context contains similarly sounding phrases (e.g., when both ``Joan'' and ``John'' are in the user's contact list) - disambiguation of the correct phrase remains challenging.

In ASR, two phrases can be very similar to each other \emph{phonetically} (i.e., in the way they sound), but be unquestionably different (e.g., ``call Joan'' and ``call John''). For a neural ASR model, the learned representations for these names might be very similar, leading the model to predict the wrong one. This problem is especially challenging for E2E ASR models presented with \emph{rare} and \emph{difficult to spell} words - as the model might not observe these words at all during training, and will thus fail to spell them correctly at test time.

In this work, we present an approach for \emph{focusing} on rare phrases during training and teaching the model to distinguish these phrases from phonetically similar ones. Using this approach, we expect the model to learn more robust representations that will help it to perform better at test time. The main ideas are to (i) detect difficult to transcribe and rare words in the input utterance as the target of focus, and (ii) train harder on these words by providing the model with difficult negative examples. 

For detection of the phrases to focus on ((i)), we find that \emph{proper nouns} (also tagged as ``NNP'' \cite{marcus1993building}) are the general category of phrases that are rare and usually more difficult to transcribe, while being relatively easy to detect \cite{huang2015bidirectional, andor2016globally, ratnaparkhi1996maximum}. For training harder on these phrases ((ii)), we extract phonetically similar alternative phrases, and feed those to the model as negative examples. 

This approach can be thought of as \emph{data augmentation} for speech. While data augmentation is usually used in machine learning for generating mutated \emph{positive} examples \cite{krizhevsky2012imagenet, KimMisraChinEtAl17}, our approach is used for generating mutated \emph{negative} examples. We apply our approach to the recently proposed Contextualized Listen, Attend, and Spell model \cite{pundak2018deep} and show that the proposed method gives up to a $53.1\%$ relative improvement in WER across several benchmarks that represent Google's voice search contextual traffic.



%% file: background.tex
\section{Background: CLAS Model}\label{sec:background}


The Contextual Listen, Attend, and Spell (CLAS) model \cite{pundak2018deep} is an E2E ASR system based on the Listen, Attend and Spell (LAS) model \cite{Chan15}, with the key difference that it also utilizes \emph{context}.
This model was shown to be a strong all-neural approach for contextual biasing. 

The CLAS model outputs a probability distribution $P(\vec{y} | \vec{x}, \vec{z})$ over sequences of output labels $\vec{y}$ (graphemes, in this work)
conditioned both on a sequence of input audio frames $\vec{x} = (\vec{x}_1, \dots, \vec{x}_K)$ 
\emph{and} a set of \emph{bias phrases} $\vec{z} = (\vec{z}_1, \dots, \vec{z}_N)$ -
a set of word n-grams, some of which might appear in the reference transcript. 
For example, if a user says to a speech-enabled device \texttt{call joan's mobile}, the system can leverage all the contacts in the device to serve as bias phrases, hopefully attending on a contact name which is called \texttt{joan}. A special ``n/a'' symbol is used to allow the model not to use any of the bias phrases at a certain decoding step.

\begin{figure}[h!]
\centering
\includegraphics[height=4cm,keepaspectratio]{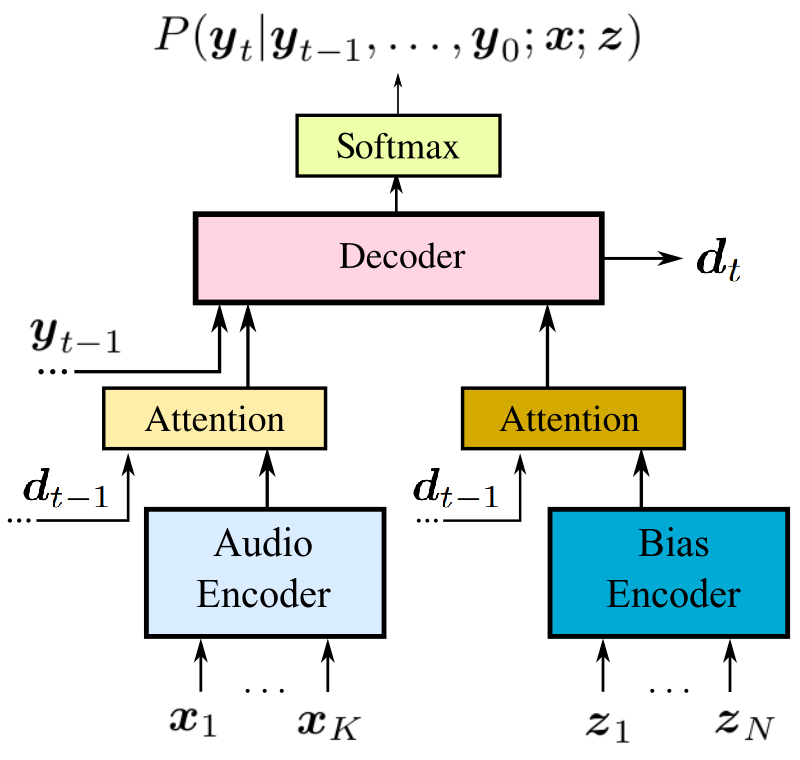}
\caption{A schematic illustration of the CLAS architecture \cite{pundak2018deep}.}
\label{fig:clas-arch}
\vspace{-2mm}
\end{figure}

CLAS uses an attention-based encoder-decoder architecture \cite{BahdanauChorowskiSerdyukEtAl16} with two encoders: an audio encoder, which computes embedding for the audio inputs $\vec{x}$, and a bias-encoder which embeds the context phrases $\vec{z}$.
At each decoding step, the decoder attends to the audio inputs \emph{and simultaneously} attends to the set of bias phrases, where the decoder state, $\vec{d_t}$ is used as attention-query.
CLAS emits its predictions a label at a time, conditioned on inputs and on previous predictions:
$P(\vec{y}_t | \vec{y}_{<t}; \vec{x} ; \vec{z})$.
\Cref{fig:clas-arch} illustrates the CLAS model. For a full description we refer the reader to \cite{pundak2018deep}.

In the original CLAS model \cite{pundak2018deep}, during training the bias phrases were randomly sampled n-grams from the reference transcript and other utterances in the training data. At test time, the bias phrases were all  \emph{from the same semantic category} (e.g., contact names). This made the \emph{test task harder than the training task}, since distinguishing between first names (such as: \texttt{joan} vs. \texttt{john}) is usually more challenging than distinguishing between random unrelated n-grams.

To close this train-test discrepancy, in this work we sample \emph{proper nouns} from the reference transcript instead of random n-grams.
Furthermore, we augment each of these proper nouns with phonetically similar alternatives.
For example, in \texttt{call joan's mobile}, we select \texttt{joan} as a bias phrase (because it is a name of a person and thus a proper noun), and phonetically similar names such as \texttt{john} and \texttt{jean} as negative examples. This approach allows to train the model on a more difficult task of distinguishing between phonetically similar names. 


%% file: our_approach.tex
\section{Training CLAS with difficult examples}\label{sec:training}
\begin{figure}[h!]
\centering
\includegraphics[width=\linewidth]{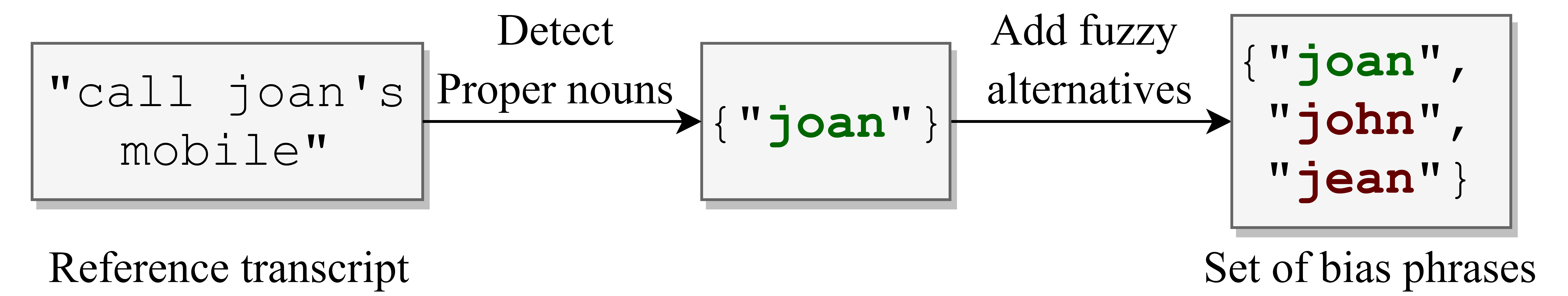}
\caption{Our approach of detecting proper nouns and augmenting them with phonetically similar alternatives, for creating sets of bias phrases for training.}
\label{fig:pipeline}
\vspace{-5mm}
\end{figure}

We now explain the major components of our approach.
Given a reference transcript, our goal is to construct a set of bias phrases. First, we extract proper nouns from the reference (\Cref{ssec:detecting}); second, we add phonetically similar (``fuzzy'') alternative phrases (\Cref{ssec:obtaining}). We then use the set of extracted proper nouns with their fuzzy alternatives as the set of bias phrases for training the model on that example. \Cref{fig:pipeline} illustrates our approach in a high-level.

\subsection{Detecting proper nouns}\label{ssec:detecting}
During training we restrict our choice of bias phrases to \emph{proper nouns}.
To achieve this, we analyze each reference transcript with 
a Part-Of-Speech (POS) tagger.
We use Google's Cloud Natural Language API \cite{google_nl_api} for POS tagging, which employs neural models similar to \cite{huang2015bidirectional, andor2016globally}.
The POS model was trained using about $1,700$ English news and web documents, containing about $500k$ tokens.

\subsection{Obtaining fuzzy alternatives}\label{ssec:obtaining}

Once we have identified a set of proper nouns from the reference transcript, we next generate fuzzy alternatives as additional bias phrases for training. We define a word n-gram $w_1$, as a \emph{fuzzy alternative} of a word n-gram $w_2$, if both are phonetically similar and co-occur often in different decoding beams of the same utterances in training data.  This process is described as follows. 

First, we start by building a fuzzy inventory that stores for each n-gram $w$ a set of fuzzy alternatives. We build the fuzzy inventory in an unsupervised manner, by utilizing an external conventional model, trained as in \cite{variani2017end}, which we use to decode a large corpus of utterances.
For each decoded utterance we obtain a set of hypotheses. We then count all the co-occurrences of n-gram pairs that appear in different hypotheses where the rest of their hypotheses are identical, and each pair is scored according to that count. For instance, ``john lemon'' will get a high score with ``john lennon'' if these can be often found in different hypotheses of the same utterances.

Next, we use the fuzzy inventory during CLAS training. Specifically, given a word n-gram representing a bias phrase, fuzzy alternatives are selected from the fuzzy inventory and are sorted according to the co-occurrence score. We further filter the top ranking candidates by keeping only those that are phonetically similar to our target n-gram, where phonetic similarity is measured using the Hixon metric \cite{hixon2011phonemic}.
This process of selecting fuzzy bias phrase alternatives is done as part of the data preparation phase.

%% file: experiments.tex
\section{Experiments}\label{sec:experiments}

Our evaluation aims to answer \emph{whether training a model using our proposed approach improves the results over CLAS}, which is a strong all-neural baseline approach for contextual biasing~\cite{pundak2018deep}. We use the same architecture, training and test corpora as \cite{pundak2018deep}. 
The baseline CLAS model is fed with random n-grams from the reference transcript and other random n-grams from the rest of the training data as bias phrases. In contrast, in our proposed approach we select \emph{proper nouns} from the reference transcript (as detailed in \Cref{ssec:detecting}) and phrases that are \emph{phonetically similar} to the selected proper nouns (as detailed in \Cref{ssec:obtaining}). At test time, all of the models are presented with the same bias phrases for each example.

\subsection{Experimental Setup}\label{ssec:setup}
Our training setup is very close to \cite{pundak2018deep}.
Our training set includes $\sim$25,000 hours of audio consisting of
$33$ million English utterances, anonymized and hand-transcribed, 
representative of Google's voice search traffic.
This data set is augmented by artificially corrupting clean utterances using a
room simulator, adding varying degrees of noise and reverberation such that the
overall SNR is between 0dB and 30dB, with an average SNR of
12dB~\cite{KimMisraChinEtAl17}.
The noise sources are from YouTube and daily life noisy environmental
recordings.
The models are trained on $8\times8$ Tensor Processing Units (TPU)
slices with global batch size of 4,096. Each training core operates on a shard-size
of $32$ utterances in each training step.

We use $80$-dimensional log-mel acoustic features computed every 10ms over a 25ms window.
Following~\cite{CC18} we stack $3$ consecutive frames and stride the stacked
frames by a factor of $3$.
The encoder's architecture consists of $10$ unidirectional LSTM layers, each with $256$ nodes.
The encoder-attention is computed over $512$ dimensions, using $4$ attention heads.
The bias-encoder consists of a single LSTM layer with $512$ nodes and the bias-attention is computed over $512$ dimensions.
Finally, the decoder consists of $4$ LSTM layers with $256$ nodes. In total, the model has about $58$ million trainable parameters and is implemented using TensorFlow~\cite{AbadiAgarwalBarhamEtAl15}.

\subsection{Test sets}\label{ssec:test-sets}


\begin{table}[h!]
\centering
  \vspace{-3mm}
  \scriptsize
    \begin{tabular}{ lrrrr }
      \toprule
      Test Set         & \makecell[cc]{Number of\\ utterances} & \makecell[cc]{Avg number of\\ bias phrases}  \\
      \midrule
      \emph{Songs}           & 15k   & 303         \\
      \emph{Contacts}        & 15k   & 75          \\
      \emph{Talk-To}         & 4k    & 3,255       \\
      \emph{Voice Search}    & 14k   & -           \\
      \emph{Dictation}       & 15k   & -           \\
      \bottomrule
    \end{tabular}
\caption{The test sets. Voice Search and Dictation do not contain any bias phrases and are thus testing a pure ASR task.} 
\label{table:test-sets}
\end{table}

\Cref{table:test-sets} presents a summary of the test sets. Specifically, each of \emph{Songs}, \emph{Contacts}, and \emph{Talk-To} contains utterances with a distinct set of bias phrases which
vary from four phrases up to more than three thousand phrases.
The test examples were artificially generated using a \emph{Parallel WaveNet} \cite{oord2017parallel} text-to-speech (TTS) engine, and
corrupted with noise similar to the training data \cite{KimMisraChinEtAl17}. We refer to the \emph{Songs}, \emph{Contacts} and \emph{Talk-To} as ``contextualized test sets'', as these contain context relevant for recognition (e.g., contact names, songs names and chatbot names). For more information regarding these test sets see \cite{pundak2018deep}. The \emph{Voice Search} and \emph{Dictation} test sets are composed of anonymized and hand-transcribed utterances, sampled from real traffic. They do not include contextual information, and are thus referred to as the ``context-free test sets''. We use the context-free test sets to evaluate the models on a pure ASR task (\Cref{ssec:results-unbiased-sets}). 

\begin{table}[t]
\centering
  \vspace{-3mm}
  \scriptsize
    \begin{tabular}{ lllll }
      \toprule
      Test Set         & \makecell[cc]{Vanilla\\CLAS} & CLAS+NNP & CLAS+fuzzy & \makecell[cc]{CLAS\\NNP+fuzzy} \\ 
      \midrule
      \emph{Songs}      & \,\,9.8           & \,\,6.7 (31.6\%)          & 10.4            & \,\,\textbf{5.4} (44.9\%)  \\
      \emph{Contacts}   & 11.3          & \,\,6.1 (46.0\%)          & 16.5           & \,\,\textbf{5.3} (53.1\%)  \\
      \emph{Talk-To}    & 15.2          & 14.8 (2.6\%)         & \textbf{11.1} (27.0\%)          & 11.3 (25.7\%) \\
      \bottomrule
    \end{tabular}
\caption{WER of the compared models on the biasing task. The relative improvement over Vanilla CLAS appears in parentheses (if the model improves).}
\label{table:bias_test_sets_results}
\vspace{-5mm}
\end{table}

\subsection{Compared models}\label{ssec:compared-models}
We compare several models that have the same architecture but were trained with a different mix of bias phrases for each example: 

\para{``Vanilla'' CLAS} - was trained with random n-grams from the reference transcript and other unrelated random n-grams from the rest of the training data as bias phrases (as in \cite{pundak2018deep}).

\para{CLAS+NNP} - was trained with proper nouns as bias phrases, both from the reference transcript and other random proper nouns from the rest of the training data.

\para{CLAS+fuzzy} - was trained with random n-grams from the reference transcript, their fuzzy alternatives, and other random n-grams from the rest of the training data and their fuzzy alternative.

\para{CLAS NNP+fuzzy} - combined both of the approaches: the bias phrases contained proper nouns from the reference transcript, their fuzzy alternatives, and other random proper nouns.

The overall max size of the set of bias phrases in each training scheme was the same ($64$), the only differences between the compared models were (i) which bias phrases were selected from the reference transcript - random n-gram or proper nouns; and (ii) whether we included their fuzzy alternatives or only random phrases. In the models that use proper nouns, at most $3$ proper nouns are selected at random from each example.
In the models that use fuzzy alternatives, $3$ fuzzy alternatives are added for each source phrase. 

\subsection{Biasing task}\label{results-biased-sets}
\Cref{table:bias_test_sets_results} shows the results for each of the compared training schemes, across the contextualized test sets.
As shown, CLAS NNP+fuzzy decreases the WER compared to Vanilla CLAS by $44.9\%$ on Songs, $53.1\%$ on Contacts, and $25.7\%$ on Talk-To. On Songs and Contacts, using proper nouns (CLAS+NNP) contributes the most, while fuzzy (CLAS+fuzzy) degrades the result. Surprisingly, their combination (CLAS NNP+fuzzy) achieves a lower WER than each of them solely. 
On Talk-To, using proper nouns (CLAS+NNP) makes a minor improvement of $2.6\%$, fuzzing solely (CLAS+fuzzy) improves by $27\%$, and using both proper nouns and fuzzing (CLAS NNP+fuzzy) achieves almost the same WER as fuzzing only.
We find the CLAS NNP+fuzzy model to perform the best, while taking the best out of each of its combined techniques. 

\subsection{Varying the number of bias phrases}\label{sssec:varying}
\input{varying_figure}

To further understand the source of improvement, we varied the number of bias phrases in Talk-To, which has the largest number of bias phrases. We included only the correct bias phrases (the ones that were extracted from the reference) and a varying number of other distracting phrases, between $0$ and $3,255$.
As shown in \Cref{fig:varying-results}, 
CLAS+NNP achieves a low WER when presented with a small set of bias phrases (which is consistent with the superiority of CLAS+NNP on Songs and Contacts), but performs worse on the full set; in contrast, CLAS+fuzzy scales well to a large number of bias phrases and achieves the lowest WER when presented with the full set of $3,250$ bias phrases, but does not utilize small sets, even when presented only with the correct phrases; 
CLAS NNP+fuzzy takes the best out of both world: it achieves the lowest WER on small sets of bias phrases thanks to the proper nouns, and achieves almost the lowest WER (only $0.2$ higher than CLAS+fuzzy) on the full set of bias phrases thanks to the fuzzy alternatives.


\subsection{Pure ASR task}\label{ssec:results-unbiased-sets}

\begin{table}[t!]
\centering
  \vspace{-3mm}
  \scriptsize
    \begin{tabular}{lrrrr}
      \toprule
      Test Set          & \makecell[cc]{Vanilla\\CLAS} & \makecell[cc]{CLAS\\NNP+fuzzy} &  \makecell[cc]{CLAS\\NNP+fuzzy\\$\alpha_{drop}=5\%$} & \makecell[cc]{CLAS\\NNP+fuzzy\\$\alpha_{drop}=30\%$}\\
      \midrule
      \textbf{Pure ASR Tasks} &&& \\ 
      \emph{Voice Search}      & \textbf{6.4}           & 8.8  & 7.7 & 6.8\\
      \emph{Dictation}         & 5.6           & 6.2  & 5.9 & \textbf{5.5} \\
       \midrule
      \textbf{Biasing Tasks} &&& \\ 
      \emph{Songs}           & 9.8           & \textbf{5.4} &  5.5   & 6.6 \\
      \emph{Contacts}        & 11.3          & 5.3 &  \textbf{5.1}   & 5.7 \\
      \emph{Talk-To\protect\footnotemark}         & 8.9	         & \textbf{8}   &	8.1	 & 8.1 \\
      \bottomrule
    \end{tabular}
\caption{WER for the Pure-ASR tasks. Loses on the Pure-ASR task can be almost recovered with $\alpha_{drop}=30\%$ while improving on the biasing task compared to Vanilla CLAS.}
\label{table:unbias_test_sets_results}
\vspace{-0.1in}
\end{table}

We evaluate each of the trained models on the context-free test sets which do not contain any bias phrases, and thus evaluate each model on a pure ASR task. The trained models are the same as in \Cref{results-biased-sets} and only the test sets are different, since we seek for a single trained model that performs well on multiple tasks.
When evaluated on these test sets, the models are trained as usual, with bias phrases, but not presented with any of those at test time.

\Cref{table:unbias_test_sets_results} shows the results of each of the compared training schemes, across the context-free test sets.
As shown, while CLAS NNP+fuzzy significantly improves over Vanilla CLAS on the biasing task (\Cref{results-biased-sets}), the improvement comes at the cost of degrading Pure-ASR performance (\Cref{table:unbias_test_sets_results}).
To address this degradation, we introduce a parameter $\alpha_{drop}$: during training, each example is presented with \emph{no bias phrases} with probability of $\alpha_{drop}$, thus making the model train on a pure ASR task in those cases. 

As shown in \Cref{table:unbias_test_sets_results}, using $\alpha_{drop}=5\%$ improves the WER in the pure ASR task compared to CLAS NNP+fuzzy (which uses $\alpha_{drop}=0\%$), while having a negligible impact on Songs and Contacts.
Using $\alpha_{drop}=30\%$ further improves the WER on the pure ASR task, without much negative impact on the biasing tasks, while still achieving much lower WER for on biasing tasks compared to Vanilla CLAS.
Further tuning of the value of $\alpha_{drop}$ allows to tune the trade-off between succeeding on biased and unbiased ASR tasks.

\footnotetext{In the experiments described in \Cref{table:unbias_test_sets_results} we found it useful to evaluate Talk-To with bias-conditioning as described in \cite{pundak2018deep}, as otherwise the model with $\alpha_{drop}=30\%$ became too sensitive to the large number of bias phrases.}

%% file: varying_figure.tex
\definecolor{nicepurple}{HTML}{AB30C4}

\begin{figure}[h!]
\begin{tikzpicture}[scale=1.05]
	\begin{axis}[
		xlabel={Max number of distracting bias phrases},
		ylabel={\footnotesize{WER}},
		ylabel near ticks,
        legend style={at={(1,0.25)},anchor=east,font=\tiny,mark size=2pt},
        xmin=0, xmax=3250,
        ymin=2, ymax=16,
        xtick={0,1000,...,3000},
        ytick={0,2,...,18},
        grid = major,
        major grid style={dotted,black},
        width = \linewidth, height = 5cm
    ]
	
    \addplot[loosely dashed,color=brown, mark options={solid, fill=brown, draw=black}, mark=triangle*, line width=0.5pt, mark size=2pt] coordinates {
		(0,6.3)
		(10,6.3)
		(100,6.2)
		(1000,11)
		(2000,13.3)
		(3000,15)
		(3250,15.2)
	}  node[above left,pos=0.85,color=black] {\tiny{Vanilla CLAS}};
    \addlegendentry{\tiny Vanilla CLAS}
    
    \addplot[densely dotted,color=blue, mark options={solid, fill=blue, draw=black}, mark=diamond*, line width=0.5pt, mark size=2pt] coordinates {
		(0,9.8)
		(10,9.8)
		(100,9.7)
		(1000,9)
		(2000,10.4)
		(3000,10.9)
		(3250,11.1)
	} node[above right,pos=0,color=black] {\tiny{CLAS+fuzzy}};
    \addlegendentry{\tiny CLAS+fuzzy}
    
    \addplot[densely dashed,color=red, mark options={solid, fill=red, draw=black}, mark=square*, line width=0.5pt, mark size=2pt] coordinates {
		(0,3.8)
		(10,3.8)
		(100,3.9)
		(1000,9.3)
		(2000,12.6)
		(3000,13.5)
		(3250,13.9)
	} node[above,pos=0.87,color=black] {\tiny{CLAS+NNP}};
    \addlegendentry{\tiny CLAS+NNP}
    
    \addplot[color=nicepurple,mark options={fill=nicepurple, draw=black, line width=0.5pt}, line width=1pt, mark=*, mark size=2pt] coordinates {
		(0,3.7)
		(10,3.7)
		(100,4.1)
		(1000,10.6)
		(2000,11.8)
		(3000,11)
		(3250,11.3)
	} node[above,pos=0.80,color=black] {\tiny{CLAS+NNP+fuzzy}} ;
    \addlegendentry{\tiny CLAS NNP+fuzzy}

	\end{axis}

\end{tikzpicture}
\caption{CLAS NNP+fuzzy achieves the lowest WER with a small set of bias phrases, and almost the lowest WER ($0.2$ higher than CLAS+fuzzy) when presented with $3255$ bias phrases. This experiment was performed on the Talk-To set.}
\label{fig:varying-results}
\end{figure}
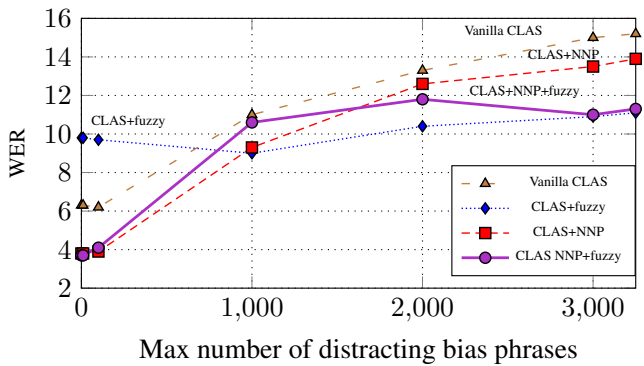 

%% file: qualitative.tex
\section{Qualitative Analysis}\label{sec:quali}
In this section we try to understand the differences in the internal representations that training with fuzzy distractors causes. We inspect the attention that the model has given to proper nouns through the decoding time steps, across different models. We use the same trained models as before, but to amplify the differences we evaluate them with one bias phrase that can be found in the transcript and $9$ fuzzy alternatives. We inspect the attention weights through the decoding steps.
We compare the CLAS+NNP model with CLAS NNP+fuzzy model, and thus inspect the effect of fuzziness. Since both of the models were trained on proper nouns, for brevity in this section we will omit the term ``NNP'' and refer to the CLAS NNP+fuzzy as ``the fuzzy model'' and to CLAS+NNP as ``the non-fuzzy model''.

In each of the following figures, the X-axis represent the decoding time steps; the Y-axis contains the top attended bias phrases by the model (the true phrase is marked with ``**''); and brighter pixels represent higher attention weights by the model per bias phrase and time step.
We classify the differences into three main categories.

\subsection{Better discrimination}\label{ssec:discrimination}
In many cases, we observed that when presented with phonetically similar phrases at test time, the fuzzy model simply \emph{captures the subtle phonetic differences better} than the non-fuzzy model. This is expressed by both a more accurate prediction and more attention on the bias phrase that actually appears in the reference transcript rather than its fuzzy alternatives. 
This affirms our hypothesis that training using fuzzy distractors makes the model discriminate phonetically similar phrases better. \Cref{fig:discriminative} shows an example.

\newcommand{\attentionwidth}{0.48\linewidth}

\begin{figure}[h!]
\centering

\centering
\begin{subfigure}{\linewidth}
\centering
\footnotesize
\textbf{True ref: \texttt{creepy carrots</bias>}}
\end{subfigure}
\\
\begin{minipage}{\linewidth}
\hspace{-2mm}
\begin{tabular}{ll}
\begin{subfigure}{\attentionwidth}
\includegraphics[width=\linewidth]{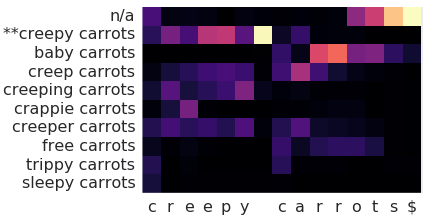}
\end{subfigure}

\begin{subfigure}{\attentionwidth}
\includegraphics[width=\linewidth]{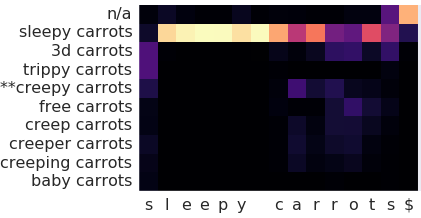}
\end{subfigure}
\\
\hspace{-2mm}
\begin{subfigure}{\attentionwidth}
\centering
\scriptsize
\fbox{Fuzzy: \texttt{\textbf{creepy} carrots</bias>}}
\end{subfigure}

\begin{subfigure}{\attentionwidth}
\centering
\scriptsize
\fbox{Non-fuzzy: \texttt{\textbf{sleepy} carrots</bias>}}
\end{subfigure}
\end{tabular}
\end{minipage}
\caption{The fuzzy model attends mostly to ``creepy carrots'' and makes a correct prediction, while the non-fuzzy model attends to ``sleepy carrots'' and predicts the wrong word ``sleepy''.}
\label{fig:discriminative}
\vspace{-5mm}
\end{figure}

\subsection{Cleaner attention distribution}\label{ssec:cleaner}
\begin{figure}[h!]
\centering
\begin{subfigure}{\linewidth}
\centering
\scriptsize
\textbf{True ref: \texttt{houston community college missouri city</bias>}}
\end{subfigure}
\\
\begin{subfigure}{\linewidth}
\includegraphics[width=\linewidth]{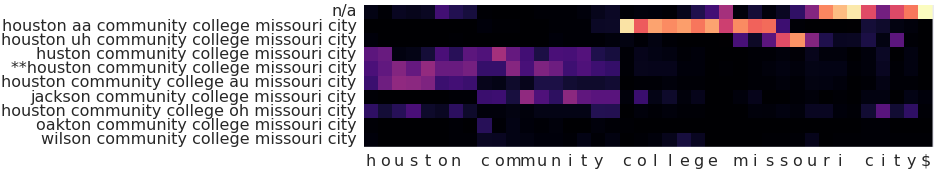}
\end{subfigure}
\\
\begin{subfigure}{\linewidth}
\centering
\scriptsize
\fbox{Fuzzy: \texttt{houston community college missouri city</bias>}}
\end{subfigure}
\\
\begin{subfigure}{\linewidth}
\includegraphics[width=\linewidth]{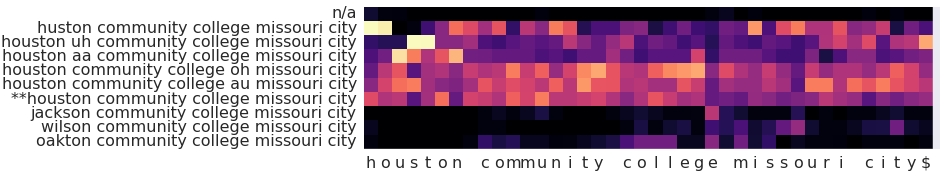}
\end{subfigure}
\\
\begin{subfigure}{\linewidth}
\centering
\scriptsize
\fbox{Non-fuzzy: \texttt{houston community college missouri city</bias>}}
\end{subfigure}

\caption{Both of the models predict the correct transcript, but the attention distribution of the fuzzy model is much cleaner and sharper.}
\label{fig:cleaner-attention}
\vspace{-5mm}
\end{figure}

We observed that even when the predictions of both models are correct, the fuzzy model usually attends more sharply and its attention distribution is much cleaner than the non-fuzzy model, which includes incorrect phrases in its attention. \Cref{fig:cleaner-attention} shows an example.

\subsection{Attention shift}\label{ssec:attention-shift}
\begin{figure}[h!]
\centering
\begin{subfigure}{\linewidth}
\centering
\scriptsize
\textbf{True ref: \texttt{...the hillcrest baptist church</bias>...}}
\end{subfigure}
\\
\begin{subfigure}{\linewidth}
\includegraphics[width=\linewidth]{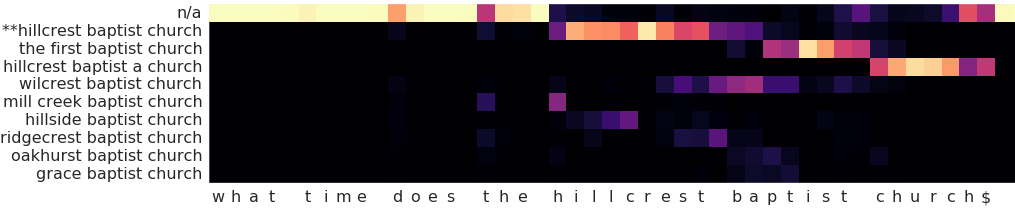}
\end{subfigure}
\\
\begin{subfigure}{\linewidth}
\centering
\scriptsize
\fbox{Fuzzy: \texttt{...the hillcrest baptist church</bias>...}}
\end{subfigure}
\\
\begin{subfigure}{\linewidth}
\includegraphics[width=\linewidth]{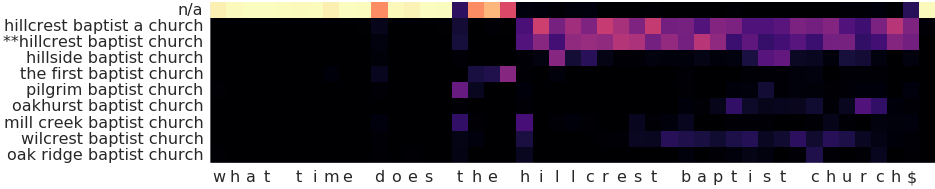}
\end{subfigure}
\\
\begin{subfigure}{\linewidth}
\centering
\scriptsize
\fbox{Non-fuzzy: \texttt{...the hillcrest baptist church</bias>...}}
\end{subfigure}

\caption{The non-fuzzy model attends to several phrases equally, while the fuzzy model shifts its attention to different phrases while decoding different words.}
\label{fig:attention-shift}
\end{figure}

When the bias phrases are composed of multiple words, we often observe an \emph{attention shift}: even though the bias phrases share some of their words, it seems that the fuzzy model has learned a ``representative'' for each bias phrase, and attends to the entire bias phrase while decoding its representative. Then, while decoding the representative of another bias phrase, the attention ``shifts'' to the other bias phrase, even if former bias phrases also contains the same word that is currently being decoded. The non-fuzzy model simply attends to several similar phrases equally while decoding the whole phrase. \Cref{fig:attention-shift} shows an example.

%% file: conclusion.tex
\section{Conclusion}
In this work we presented a general approach for training contextualized neural speech recognition models with difficult negative examples. The core idea is to detect and focus on proper nouns (``NNP'') in the reference transcript, and present the model with phonetically similar (``fuzzy'') phrases as their negative examples.

We demonstrate our approach by applying it to a speech biasing task, and show that our approach improves WER by up to $53.1\%$. By using mixed training we are able to tune the trade-off between accuracy on the biasing and the pure ASR tasks. 
We analyze the contribution of each of the components of our approach and discover interesting phenomena in the model's internal attention that are consistent with our results.
We believe that the principles presented in this paper can serve as a basis for a wide range of speech tasks. 

%% file: main.bbl
\begin{thebibliography}{10}

\bibitem{Chan15}
William Chan, Navdeep Jaitly, Quoc Le, and Oriol Vinyals,
\newblock ``Listen, attend and spell: A neural network for large vocabulary
  conversational speech recognition,''
\newblock in {\em Acoustics, Speech and Signal Processing (ICASSP), 2016 IEEE
  International Conference on}. IEEE, 2016, pp. 4960--4964.

\bibitem{Sak15}
Hasim Sak, Andrew~W. Senior, Kanishka Rao, and Fran{\c{c}}oise Beaufays,
\newblock ``Fast and {A}ccurate {R}ecurrent {N}eural {N}etwork {A}coustic
  {M}odels for {S}peech {R}ecognition,''
\newblock in {\em INTERSPEECH}, 2015.

\bibitem{SoltauLiaoSak16}
Hagen Soltau, Hank Liao, and Haşim Sak,
\newblock ``Neural speech recognizer: Acoustic-to-word lstm model for large
  vocabulary speech recognition,''
\newblock in {\em Proc. Interspeech 2017}, 2017, pp. 3707--3711.

\bibitem{BahdanauChorowskiSerdyukEtAl16}
D.~Bahdanau, J.~Chorowski, D.~Serdyuk, P.~Brakel, and Y.~Bengio,
\newblock ``End-to-end attention-based large vocabulary speech recognition,''
\newblock in {\em Proc. of ICASSP}, 2016.

\bibitem{PrabhavalkarSainathWuEtAl18}
R.~Prabhavalkar, T.~N. Sainath, Y.~Wu, P.~Nguyen, Z.~Chen, C.-C. Chiu, and
  A.~Kannan,
\newblock ``Minimum word error rate training for attention-based
  sequence-to-sequence models,''
\newblock in {\em Proc. of ICASSP}, 2018.

\bibitem{CC18}
C.-C. Chiu, T.~N. Sainath, Y.~Wu, R.~Prabhavalkar, P.~Nguyen, Z.~Chen,
  A.~Kannan, R.~J. Weiss, K.~Rao, N.~Jaitly, B.~Li, and J.~Chorowski,
\newblock ``State-of-the-art speech recognition with sequence-to-sequence
  models,''
\newblock in {\em Proc. ICASSP}, 2018.

\bibitem{AleksicAllauzenElsonEtAl15}
P.~Aleksic, C.~Allauzen, D.~Elson, A.~Kracun, D.~M. Casado, and P.~J. Moreno,
\newblock ``Improved recognition of contact names in voice commands,''
\newblock in {\em Proc. of ICASSP}, 2015, pp. 5172--5175.

\bibitem{WilliamsKannanAleksicEtAl18}
Ian Williams, Anjuli Kannan, Petar Aleksic, David Rybach, and Tara~N. Sainath,
\newblock ``Contextual speech recognition in end-to-end neural network systems
  using beam search,''
\newblock in {\em Proc. of Interspeech}, 2018.

\bibitem{pundak2018deep}
Golan Pundak, Tara~N Sainath, Rohit Prabhavalkar, Anjuli Kannan, and Ding Zhao,
\newblock ``Deep context: end-to-end contextual speech recognition,''
\newblock {\em arXiv preprint arXiv:1808.02480}, 2018.

\bibitem{marcus1993building}
Mitchell~P Marcus, Mary~Ann Marcinkiewicz, and Beatrice Santorini,
\newblock ``Building a large annotated corpus of english: The penn treebank,''
\newblock {\em Computational linguistics}, vol. 19, no. 2, pp. 313--330, 1993.

\bibitem{huang2015bidirectional}
Zhiheng Huang, Wei Xu, and Kai Yu,
\newblock ``Bidirectional lstm-crf models for sequence tagging,''
\newblock {\em arXiv preprint arXiv:1508.01991}, 2015.

\bibitem{andor2016globally}
Daniel Andor, Chris Alberti, David Weiss, Aliaksei Severyn, Alessandro Presta,
  Kuzman Ganchev, Slav Petrov, and Michael Collins,
\newblock ``Globally normalized transition-based neural networks,''
\newblock in {\em Proceedings of the 54th Annual Meeting of the Association for
  Computational Linguistics (Volume 1: Long Papers)}, 2016, vol.~1, pp.
  2442--2452.

\bibitem{ratnaparkhi1996maximum}
Adwait Ratnaparkhi,
\newblock ``A maximum entropy model for part-of-speech tagging,''
\newblock in {\em Conference on Empirical Methods in Natural Language
  Processing}, 1996.

\bibitem{krizhevsky2012imagenet}
Alex Krizhevsky, Ilya Sutskever, and Geoffrey~E Hinton,
\newblock ``Imagenet classification with deep convolutional neural networks,''
\newblock in {\em Advances in neural information processing systems}, 2012, pp.
  1097--1105.

\bibitem{KimMisraChinEtAl17}
Chanwoo Kim, Ananya Misra, Kean Chin, Thad Hughes, Arun Narayanan, Tara
  Sainath, and Michiel Bacchiani,
\newblock ``Generation of large-scale simulated utterances in virtual rooms to
  train deep-neural networks for far-field speech recognition in google home,''
\newblock in {\em Proc. of Interspeech}, 2017.

\bibitem{google_nl_api}
Google,
\newblock ``Cloud natural language,''
  \url{https://cloud.google.com/natural-language/}, 2018,
\newblock [Online].

\bibitem{variani2017end}
Ehsan Variani, Tom Bagby, Erik McDermott, and Michiel Bacchiani,
\newblock ``End-to-end training of acoustic models for large vocabulary
  continuous speech recognition with tensorflow,''
\newblock in {\em in Proc. Interspeech}, 2017.

\bibitem{hixon2011phonemic}
Ben Hixon, Eric Schneider, and Susan~L Epstein,
\newblock ``Phonemic similarity metrics to compare pronunciation methods,''
\newblock in {\em Twelfth Annual Conference of the International Speech
  Communication Association}, 2011.

\bibitem{AbadiAgarwalBarhamEtAl15}
M.~{Abadi et al.},
\newblock ``{TensorFlow: Large-Scale Machine Learning on Heterogeneous
  Distributed Systems},'' {Available online:
  http://download.tensorflow.org/paper/whitepaper2015.pdf}, 2015.

\bibitem{oord2017parallel}
Aaron van~den Oord, Yazhe Li, Igor Babuschkin, Karen Simonyan, Oriol Vinyals,
  Koray Kavukcuoglu, George van~den Driessche, Edward Lockhart, Luis~C Cobo,
  Florian Stimberg, et~al.,
\newblock ``Parallel wavenet: Fast high-fidelity speech synthesis,''
\newblock {\em arXiv preprint arXiv:1711.10433}, 2017.

\end{thebibliography}
